\documentclass[11pt]{article}
\usepackage[normalem]{ulem}
\pdfoutput=1 % if you are submitting a pdflatex (i.e. if you have

\usepackage{jheppub} % for details on the use of the package, please
\usepackage{url}
\usepackage{caption}
\usepackage{subcaption}
\usepackage{extarrows}

\usepackage[latin9]{inputenc}
\usepackage{float}
\usepackage{amsmath}
\usepackage{amssymb}
\usepackage{graphicx}
\usepackage{esint}
\usepackage{hyperref}
\usepackage{comment}
\usepackage{color}
\usepackage{microtype}
\usepackage{cleveref}
\usepackage{breakurl}
\usepackage{bbm}
\newcommand{\be}{\begin{equation}}
\newcommand{\ee}{\end{equation}}
\newcommand{\ben}{\begin{displaymath}}
\newcommand{\een}{\end{displaymath}}
\newcommand{\bea}{\begin{eqnarray}}
\newcommand{\eea}{\end{eqnarray}}
\def\K{K{\"a}hler }
   \newcommand{\rf}[1]{(\ref{#1})}
\newcommand{\vp}{\varphi}

\def\be{\begin{equation}}
\def\ee{\end{equation}}
\def\bea{\begin{eqnarray}}
\def\eea{\end{eqnarray}}
\def\ba{\begin{array}}
\def\ea{\end{array}}
\def\bit{\begin{itemize}}
\def\eit{\end{itemize}}

\def\a{\alpha}

\def\vp{\varphi}

 \makeatletter

\allowdisplaybreaks

\makeatother

\makeatletter
\DeclareRobustCommand{\rcite}[1]{%
  \rcite@aux#1,\@nil{#1}%
}
\def\rcite@aux#1,#2\@nil#3{%
  \if\relax#2\relax
    Ref.~\cite{#3}%
  \else
    Refs.~\cite{#3}%
  \fi
}
\makeatother

\hypersetup{
    colorlinks = true,
    citecolor = {blue},
    linkcolor = {blue},
    urlcolor = {blue},
}

\def\eqn#1{eq.~\eqref{#1}}

\def\rcite#1{ref.~\cite{#1}}

 \title{\rm { \bf   Quintessential  {\boldmath $\a$}-attractors, updated}
}

\author[a]{Renata Kallosh,}
\author[a]{Andrei Linde,}
\author[a]{Marina Shmakova,}
\author[b]{Yusuke Yamada}
\affiliation[a]{Leinweber Institute for Theoretical Physics at Stanford, 382 Via Pueblo, Stanford, CA 94305, USA}
\affiliation[b]{Cosmology, Gravity, and Astroparticle Physics Group, Center for Theoretical Physics of the Universe,
Institute for Basic Science (IBS), Daejeon, 34126, Korea}
\emailAdd{kallosh@stanford.edu}
\emailAdd{alinde@stanford.edu}
\emailAdd{marina.shmakova@wvm.edu}
\emailAdd{yamada@ibs.re.kr}

\abstract{ Quintessential $\a$-attractor  models of single-field inflation and evolving dark energy were constructed about a decade ago. Recently, it was pointed out that some of them might be disfavored due to dark-radiation constraints on gravitational waves and the higher values of $n_s$ favored by ACT. Here we present a class of updated quintessential $\a$-attractor models in which a single field describes both inflation and evolving dark energy, yields higher values of $n_s$, and admits reheating scenarios consistent with the dark-radiation bound on $\Delta N_{\rm eff}$. Depending on the value of the cosmological constant $\Lambda$, these models interpolate between $\Lambda$CDM with $\Lambda > 0$ (future dS universe), dynamical dark energy with $\Lambda = 0$ (future Minkowski universe), and dynamical dark energy with $\Lambda < 0$ (future cosmological collapse). We also study quintessential $\a$-attractor models based on an axion-inflaton complex scalar field with hyperbolic geometry, which describe inflation and dark energy of a ``phantom illusion'' type compatible with DESI DR2.}

\begin{document}
\maketitle

% \tableofcontents{}
 
\parskip 7 pt

%\newpage
\section{Introduction} 
We study a class of general quintessential $ \alpha$-attractor models for inflation and dark energy with an exponential potential, with future de Sitter and Minkowski states. These models were introduced about a decade ago in \cite{Akrami:2017cir,Dimopoulos:2017zvq,Akrami:2020zxw}.

 At that time, there was no evidence against a simple cosmological constant (CC) leading to a future de Sitter universe. However,  DESI DR2 results \cite{DESI:2025zgx} favor dynamical dark energy. Still, some potential methodological concerns with respect to the DESI preference for dynamical dark energy over $\Lambda$CDM have been found in \cite{Ong:2025utx,Afroz:2025iwo,Ong:2026tta,Garcia-Garcia:2026nzy}.

New data from  Euclid,  Vera Rubin Observatory,  Nancy Grace Roman Space Telescope, and the full DESI and DES programs
 will clarify the status of dynamical dark energy. In this paper, we will take the point of view that some form of dynamical dark energy is possible and will look for models with an equation of state $w(z)$ that interpolate between $\Lambda$CDM and DESI data on dark energy.
We do not know whether Euclid, Rubin, and other experiments will support the ``phantom crossing'' results of DESI or will move us back to $\Lambda$CDM and CC.  In the long term,  a combination of all new observational data will tell us more.

For now, we will consider two possibilities:

1. Future observations may point towards evolving dark energy in combination with a cosmological constant. This matches the string-theory landscape scenario, where the cosmological constant can take a wide range of values, with either sign.  If $\Lambda$ is positive, the universe in the future will gradually approach a dS state \cite{Akrami:2017cir}. If $\Lambda = 0$, the universe will evolve into a Minkowski space \cite{Akrami:2017cir,Dimopoulos:2017zvq}. And if $\Lambda < 0$, the universe will eventually collapse  \cite{Kallosh:2002gf, Kallosh:2002gg, Kallosh:2003mt, Gutperle:2003kc, Kallosh:2003bq}. In this paper, we will consider all these possibilities in the context of the quintessential $\alpha$-attractors.

2. If future observations support DESI findings, including ``phantom crossing'' results, we will focus on a class of quintessential $\a$-attractor inflaton-axion models, where we use the fact that these models are based on hyperbolic geometry with a complex scalar. We will consider the embedding of the 
two-field quintessence model \cite{Toomey:2025yuy,Chudaykin:2026amr}, which does not violate the null energy condition but supports DESI results,  into quintessential $\a$-attractor inflaton-axion models. These models will be ready if future experiments support the dynamical dark energy favored by DESI.

Recently, when CMB data from Planck and BICEP/Keck were combined with the latest ACT and SPT data \cite{AtacamaCosmologyTelescope:2025blo,SPT-3G:2025bzu}, the value of $n_{s}$ increased to $n_{s}=0.9682\pm 0.0032$ \cite{Balkenhol:2025wms}. When one also takes into account the recent DESI DR2 data \cite{DESI:2025zgx,AtacamaCosmologyTelescope:2025blo,SPT-3G:2025bzu,Balkenhol:2025wms}, the spectral index obtained in joint CMB + DESI fits is $n_{s}=0.9728\pm0.0029$, which is in tension with predictions of the simplest $\alpha$-attractor models  \cite{SPT-3G:2025bzu,Balkenhol:2025wms}.  There are some concerns with respect to this last result because the CMB data and DESI DR2 data are in significant tension with each other \cite{Ferreira:2025lrd}, but it is prudent to consider the consequences of the possible increase of $n_{s}$. 

This issue will be addressed in this paper, in the context of the recently described waterfall-modulated $\a$-attractors   \cite{Kallosh:2026kfx} originating from the hybrid $\a$-attractors \cite{Kallosh:2022ggf}  and the recent proposals~\cite{Zhang:2026ivx,Yuan:2026xcg}. This choice allows a flexible value of $n_s$ defined by the waterfall properties and its position on the plateau, which leads to modulation or premature termination of inflation \cite{Kallosh:2026kfx}.

\section{The general single-field Exp quintessential $\a$-attractor models}\label{Sec:Exp}

One of the quintessential $\a$-attractor models in \cite{Akrami:2017cir}, expressed in terms of the hyperbolic half-plane  variables with the kinetic term $-3\a{\partial T \partial \bar T\over (T+\bar T)^2}$, has the following potential 

\begin{equation}
    V_{\rm exp}(\vp)
    =M^2\left [ e^{-2g \left(  {T+\bar T\over 2+T+\bar T}\right)}-  e^{-2g}\right]
 +\Lambda \, .
    \label{exp}
\end{equation}
where
$T= e^{-\sqrt{2\over 3\a} \vp} +i\theta$. In a supergravity version of this model, we may need a stabilizer term that forces the axion $\theta$ to remain constant, so that only a single field $\vp$ evolves.

\begin{figure}[H]
\vskip -0.5cm
\centering
\IfFileExists{DES_EXP_V_full.png}{%
\includegraphics[width=0.42\textwidth]{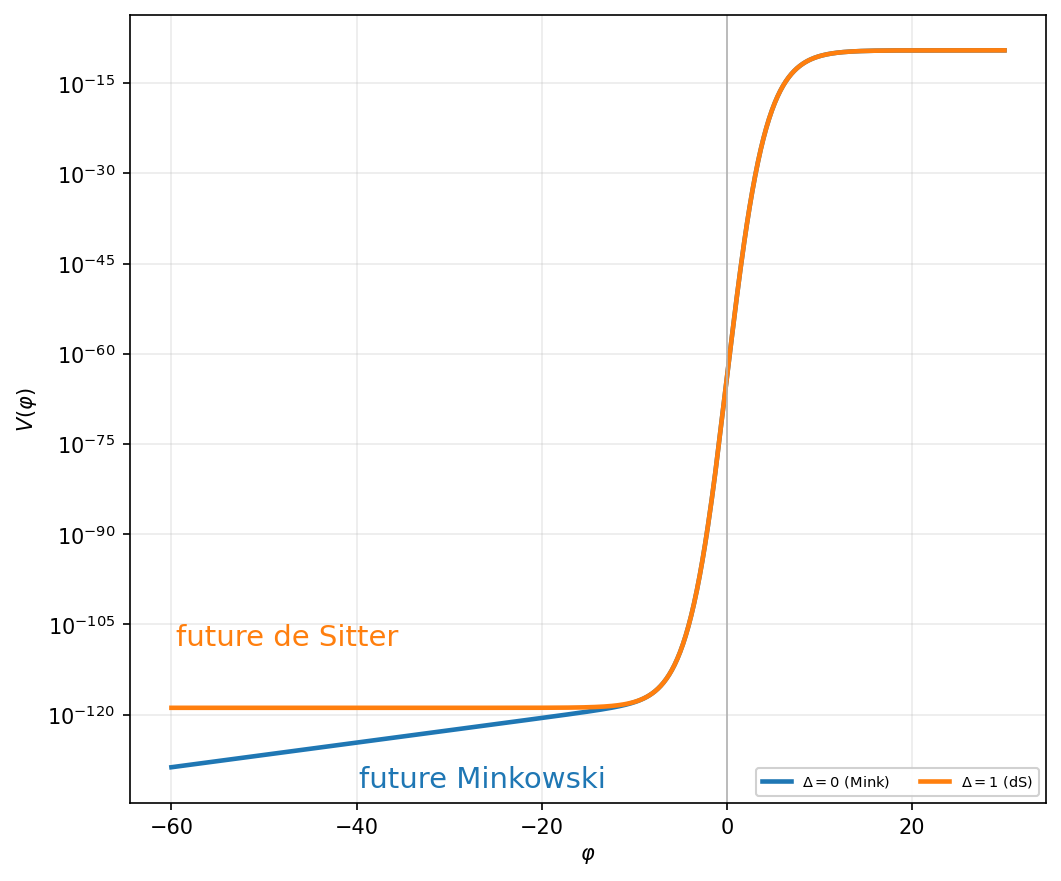}%
}{\fbox{\parbox[c][2.2in][c]{0.42\textwidth}{\centering Placeholder:
\texttt{DES\_EXP\_V\_full.png}}}}\hspace{2mm}%
\IfFileExists{logV_family_delta_DS_alpha_3_g.png}{%
\includegraphics[width=0.54\textwidth]{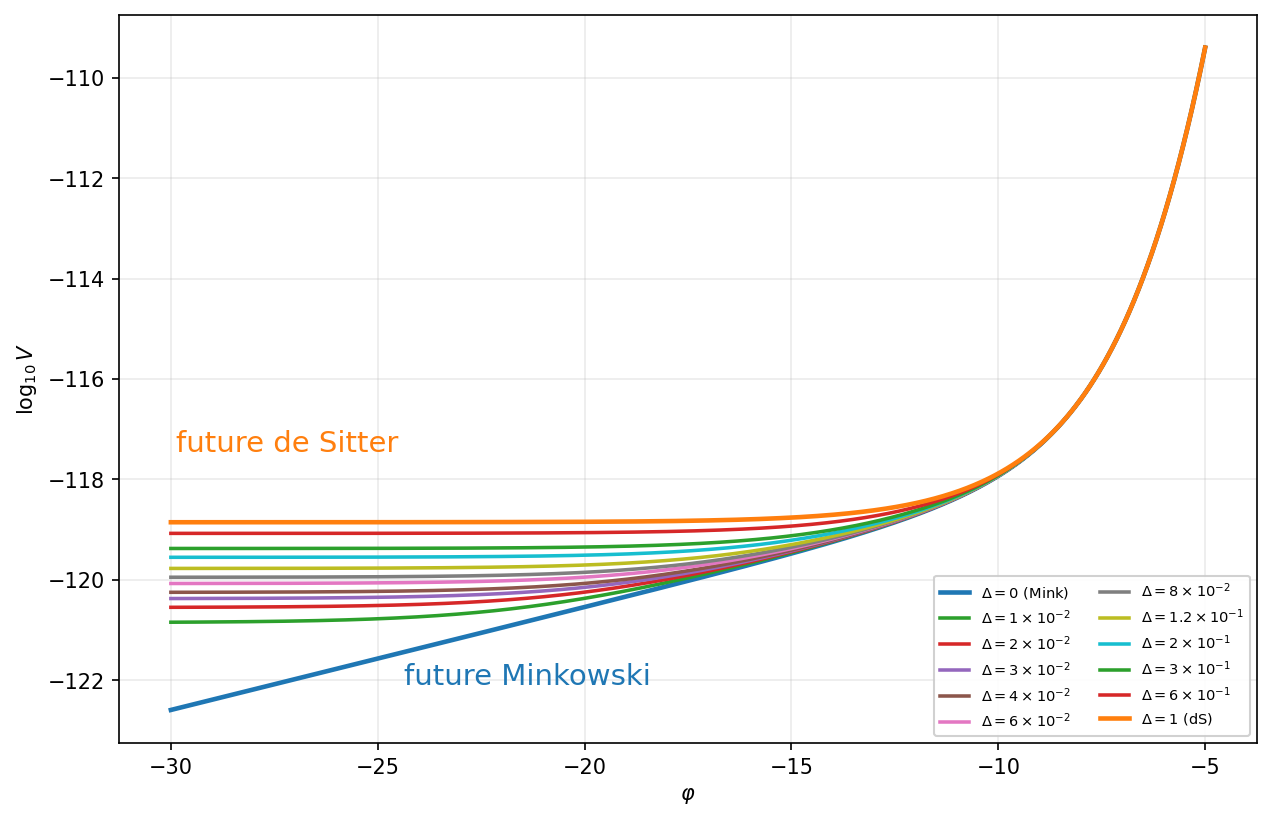}%
}{\fbox{\parbox[c][2.2in][c]{0.5\textwidth}{\centering Placeholder:
\texttt{logV\_family\_delta\_DS\_alpha\_3\_g.png}}}}
\caption{\footnotesize Exponential potential of Eq.~\eqref{exp} for models with various $\Lambda$,
from the orange line with a de~Sitter future ($\Lambda=M^2e^{-2g}$) to the blue line with a
Minkowski future ($\Lambda=0$). The left panel, from \cite{Akrami:2017cir}, shows the Exp~I
(orange) and Exp~II (blue) models with future de~Sitter and Minkowski universes,
respectively. The right panel shows the general Exp models with $\Lambda=\Delta\,M^2e^{-2g}$
interpolating between Exp~I and Exp~II, $1\geq\Delta\geq0$, with   $\Lambda$
decreasing from $\Lambda=M^2e^{-2g}$ to $\Lambda=0$ (future Minkowski universe).}
\label{InterPot}
\end{figure}

For $T=\bar T$, the potential is\footnote{ An example of a single-field quintessential $\a$-attractor model was studied in \cite{Borys:2026sna}. It was observed there that at $\a={\cal O}(1)$ the model is in good agreement with DESI.} 
\begin{equation}
    V_{\rm exp}(\vp)
    =M^2\left [ e^{g \left( \tanh\frac{\vp}{\sqrt{6\alpha}}-1\right)}-  e^{-2g}\right]
 +\Lambda \, .
    \label{exp}
\end{equation}
We can also present  it in the form
\begin{equation}
    V_{\rm exp}(\vp)
    =M^2e^{-2g}\left [ e^{g \left( \tanh\frac{\vp}{\sqrt{6\alpha}}+1\right)}-  1\right]
 +\Lambda \, .
    \label{expA}
\end{equation}
Past studies mostly considered Exp II (a future flat Minkowski universe with vanishing CC), first proposed in~\cite{Dimopoulos:2017zvq}, and Exp I (a future de Sitter universe with positive CC) proposed in \cite{Akrami:2017cir}.
The Exp II model  has 
\be
\Lambda_{{\rm{Exp}\,  II}}  =0 \,. \ee
The Exp I model  has
\be
\Lambda_{\rm{Exp}\,  I}= M^2 e^{-2g} \, .\ee
These two potentials, Exp I and Exp II,  are plotted in Fig. 10 of \cite{Akrami:2017cir}. We also show them here in the left panel of Fig. \ref{InterPot}.

\

\

% NEW PLOTS FOR V(phi) 
 
 %\ms{Updated plots for potentials }

\begin{figure}[H]
\centering
\IfFileExists{DES_SUGRA_wz_delta_DS_a3_s.png}{%
\includegraphics[width=0.6\textwidth]{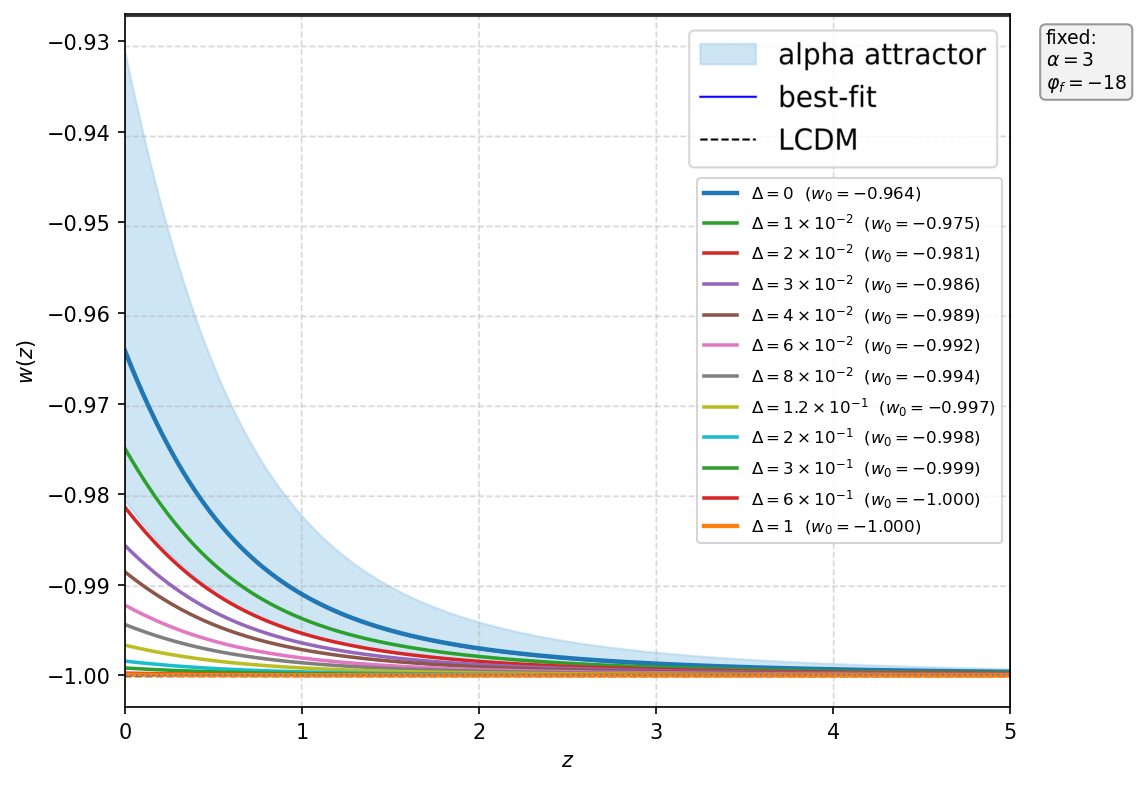}%
}{\fbox{\parbox[c][2.4in][c]{0.72\textwidth}{\centering Placeholder:
\texttt{figures/logV\_family\_k\_alpha\_3.png} --- $\log_{10}V$ vs $\varphi$ for the Exp
family, $\Delta=0\!\to\!1$.}}}
\caption{\footnotesize Dark energy equation of state $w(z)$ for models in Eq.~\eqref{exp} and the potentials in the right panel of Fig. \ref{InterPot} with  $\Lambda = \Delta M^2 e^{-2g}$. The lowest horizontal orange line is the Exp I model with $\Delta=1$; it already has $w(z)=-1$ at present, corresponding to the $\Lambda$CDM case. The upper blue curve represents an Exp II model, the best fit to DESI data in this class of models. Interpolating models between $\Lambda$CDM and DESI are those with $1 > \Delta  >0$. }
\label{Inter}
\end{figure}

Here we introduce models interpolating between Exp I and Exp II in Eq. \rf{exp}.
In these models, the cosmological constant is non-negative  
\be
 \Lambda = \Delta M^2 e^{-2g} \geq 0\ , \qquad \Delta= \{ 0,  \,  1\cdot  10^{-2} , \, 2 \cdot  10^{-2} , \, \dots  , \, 3 \cdot 10^{-1} , \, 6 \cdot  10^{-1} , \, 1\} \ , \label{CCflex}\ee
and interpolates between zero and its maximal value
\be
0\leq  \Lambda \leq M^2 e^{-2g}  
\ee
The boundary case with $ \Delta=0, \, \Lambda =0$ is Exp II, and the boundary case with $ \Delta=1, \, \Lambda =M^2 e^{-2g}$ is Exp I.
Here, in the right panel of Fig. \ref{InterPot}, we present a plot for more general values of $\Lambda$, interpolating between  Exp I and Exp II, i.e., between a future de Sitter universe with maximal $\Lambda$ in these models and a future Minkowski universe.  The models in the right panel of Fig. \ref{InterPot}  are ready for future data on dynamical dark energy, especially if they suggest a deviation from DESI towards $\Lambda$CDM.

In addition, we can include potentials with future anti-de Sitter universes, with negative $\Delta$.
These universes have a negative CC in the future. They will collapse sometime in the future; see, for example, the discussion of the fate of the universe in \cite{Kallosh:2002gf,Kallosh:2002gg,Kallosh:2003bq}.

The Exp II model, originally proposed in \cite{Dimopoulos:2017zvq} and studied in detail in \cite{Akrami:2017cir,Zhumabek:2023wka},
has attracted some attention as a candidate for a quintessential $\a$-attractor model relevant to  DESI data~\cite{Alestas:2024eic,Jing:2026ymp}. However, more recently, it was recognized that this model conflicts with dark radiation constraints due to graviton overproduction and the higher $n_s$ value from ACT + DESI \cite{Jing:2026ymp}.  In Sec. \ref{Sec:wf}, we explain how to update quintessential $\a$-attractor models to increase $n_s$ during inflation without relying on prolonged kination. Later, in Sec. \ref{Sec: rehe}, we explain how to avoid overproduction of dark radiation, including gravitons and two light scalars. We will discuss the fate of the universe in quintessential $\a$-attractor models in Sec.~\ref{Sec:fate} .

\section{Waterfall insertion in general Exp models }\label{Sec:wf}
The updated quintessential $\a$-attractor models were introduced in \cite{Kallosh:2026kfx,Chudaykin:2026amr} using the Exp II model as an example. Here we study a more general class of models that interpolate between de Sitter and Minkowski future universes, defined in \eqn{exp}. The updated version of these  quintessential $\a$-attractor models has the form   \begin{equation}
    V_{\rm upd}(\vp)=V_{\rm Exp}(\vp)\, \hat V(\vp)=
  \left (M^2 \left[ e^{g \left( \tanh \frac{\vp}{\sqrt{6\alpha}} -1\right)}-  e^{-2g}\right] 
 +\Lambda\right ) \, \hat V(\vp) \, ,
    \label{eq:v-new-step-clean}
\end{equation}
where
\begin{equation}
    \hat V(\vp)
    =\frac{1+\gamma_s\tanh \frac{\vp-\vp_c}{\Delta \vp} }{1+\gamma_s},
    \label{eq:v-step-clean}
\end{equation}
 We take the constant parameter $\gamma_s$ in the range $0<\gamma_s<1$, while $\varphi_c$ and $\Delta\varphi$ are also constants.
 
\begin{figure}[H]
\vskip 0.3cm 
\centering
		 \includegraphics[width=0.6\textwidth]{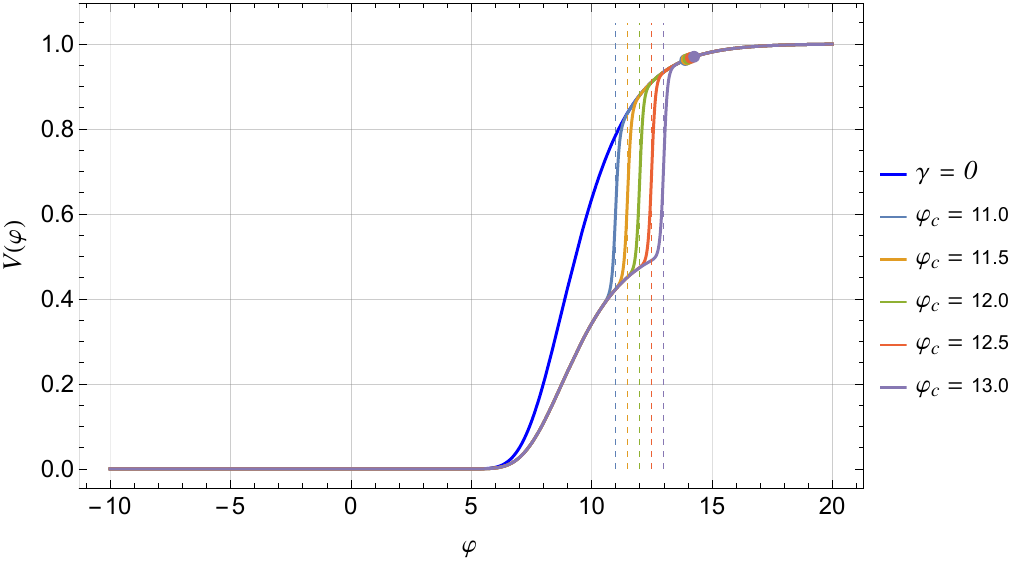}
        \caption{\footnotesize  The blue curve describes the original quintessential potential, with $\gamma_s=0$. The waterfall-modulated quintessential potentials with $\a=5/3$, $\gamma_s=0.3$,  $\Delta \vp= 0.15$  modify the inflationary evolution near the waterfall, which leads to an increase in $n_s$  in the range from $0.963$ to $ 0.9735$ for the physical number of e-foldings $N_*=55$.} 
        \label{PotQ}
\end{figure}

For example, in \cite{Kallosh:2026kfx} we examined various parameter choices defining $\hat{V}(\varphi)$. For the cases with $\a=5/3$, $\gamma_s=0.3$,  $\Delta \vp= 0.15$ and $\a=1, \gamma_s=0.9, \Delta \vp=0.612$, we plotted our potentials and presented the values of $n_s, r$ in each case. The set of potentials in the first case with various choices of $\vp_c$ is shown here in Fig. \ref{PotQ}.

As in the case studied in \cite{Kallosh:2026kfx,Chudaykin:2026amr}, one can show that the effect of the waterfall on the inflationary plateau, as shown in Fig. 5 in \cite{Chudaykin:2026amr} and explained in more detail in Sec. 4 in   \cite{Kallosh:2026kfx}, does not affect the dark energy stage of the quintessential $\a$-attractor model. 
A detailed study of the dark energy stage in the original Exp II model was performed in  \cite{Dimopoulos:2017zvq,Akrami:2017cir,Zhumabek:2023wka,Alestas:2024eic,Jing:2026ymp}. In all cases,  the initial values of the ``thawing'' quintessence field  were taken  in the range 
\be
-35 \leq \vp_i\leq -7\,,
\ee
and the results did not change much for $\vp_i $ below $-17$. For the waterfall parameter values $\gamma_s, \Delta\vp, \vp_c$ used in \cite{Kallosh:2026kfx} to increase $n_s$, it is easy to check that the value of the waterfall during inflation does not affect the dark energy equation of state $w(z)$. Namely, all the cases we study have the property that 

\begin{equation}
\frac{\vp_i -\vp_c}{\Delta \vp}\ll -1\, , \qquad    \hat V(\vp)|_{\vp \to -\infty}  \to \frac{1-\gamma_s}{1+\gamma_s}\equiv C_{\rm wf}\, .
\label{noeffect}
\end{equation}
As we have normalized the inflationary region as $\hat V\to 1$, the overall constant $C_{\rm wf}$ appears in the dark energy part.

The value of the dark energy plateau is defined by $\Delta M^2 e^{-2g}$ without the waterfall. With the waterfall, we can define $g(\gamma_s) =  g + {1\over 2} \log C_{\rm wf}$. In our example with $\gamma_s=0.3$ we find $-{1\over 2} \log C_{\rm wf} \approx 0.3$. The value of the parameter $g$ is not known precisely; it is of order $ g \approx \ln {H_{ \rm inf}\over 
H_{\rm DE}}
$. It might be in the range $120-128$ and is often taken to be $\approx 125$, so the change in $g$ caused by the waterfall, of order $  \sim 1$, is not important and, in any case, can be absorbed in the dependence of $g$ on $\gamma_s$.

In these updated models, with the waterfall $\hat V(\vp)$  insertion, the observables $n_s, r$  are given by 
the following expressions
\be
\label{pred}
n_{s} \approx  1-{2\over N_{c}} \ , \qquad r \approx {12\alpha\over N_{c}^{2}} \ .
\ee 
Here $N_c$ is the effective e-fold number induced by the waterfall factor and should be distinguished from $N_*$, the physical
number of e-folds between horizon exit of the pivot scale and the end of inflation. In the notation of~\cite{Kallosh:2026kfx}, $N_c=N_*+\Delta N$.

We will also show examples of $w(z)$ for updated models, with and without the waterfall, that confirm the qualitative analysis in this subsection.

\section{Dynamical dark energy evolution }\label{Sec:evol}

Our interpolating quintessential $\a$-attractor models depend on parameters $M^2,\, g,\, \a$ and $ \Delta$, and on the value of $\vp_f$, which is the frozen initial value of $\vp$ during dark-energy evolution. The initial velocity of the field $\vp$ at the start of dark-energy evolution in this model can be set to zero because of the high Hubble friction at early times.

According to \cite{Akrami:2017cir,Dimopoulos:2017zvq,Akrami:2020zxw,Zhumabek:2023wka},
 $M^2=10^{-10}\alpha$ in this model,  and we take into account that dark energy is expected to be $V_{\rm today} \sim 10^{-120}$.  
There is a
 price to pay for having one plateau of the model for the early universe at about
$10^{-10} $ in Planck density units, and another one for the current and future acceleration at
about $10^{-120}$. Therefore it was proposed in \cite{Akrami:2017cir} that  $g$  can be  a
parameter that is determined observationally,
\be
 g \approx \ln {H_{\rm inf}\over 
H_{\rm DE}} \, .
\ee
Since $M^2e^{-2g}$ sets the scale of the present
dark-energy density, one has $g\simeq125$  for the exponential 
inflation--dark-energy hierarchy. For each choice of  $(\alpha, \vp_f)$ one can  tune $g$ by
bisection so that the present dark-energy fraction is $\Omega_{\varphi,0}=0.7$
with $M^2=10^{-10}\alpha$.  This reproduces the range
$g\approx120$--$128$ reported  in \cite{Akrami:2017cir,Zhumabek:2023wka}.  
At $\vp_f=-18$ the representative values of $g$ are  $g= 124.11, \, 124.70, \, 125.17, \, 125.87$ for  $\a=5/3, \, 2, \, 7/3, \, 3$, respectively.

For a given choice of the quintessential potential $V(\vp)$, we follow the setup in \cite{Zhumabek:2023wka}, which is an application of the standard methods used in studies of dynamical dark energy in \cite{Copeland:2006wr}.
 The scalar-field evolution is written in terms of the autonomous variables
\begin{equation}
    x = \frac{\vp'}{\sqrt{6}},
    \qquad
    y = \frac{\sqrt{V(\vp)}}{\sqrt{3}H},
    \qquad
    {}' \equiv \frac{d}{d\ln a}.
\end{equation}
The scalar-field density fraction and equation of state are
\begin{equation}
    \Omega_\vp = x^2+y^2,
\end{equation}
\begin{equation}
    w_\vp = \frac{x^2-y^2}{x^2+y^2}.
\end{equation}

The numerical system that we evolve is
\begin{equation}
    \vp' = \sqrt{6}x,
\end{equation}
\begin{equation}
    x'
    =
    -3x
    +\sqrt{\frac32}\lambda(\vp)y^2
    +\frac32x
    \left[2x^2+\gamma(1-x^2-y^2)\right],
    \label{eq:xprime}
\end{equation}
\begin{equation}
    y'
    =
    -\sqrt{\frac32}\lambda(\vp)xy
    +\frac32y
    \left[2x^2+\gamma(1-x^2-y^2)\right],
    \label{eq:yprime}
\end{equation}
where
\begin{equation}
    \lambda(\vp) = -\frac{V_{,\vp}}{V}\, ,  \qquad 
    \gamma(a)  =  1+ \frac{1}{3}\frac{1}{1+(\Omega_{m0}/\Omega_{r0})a}.
\end{equation}
We refer the reader to \cite{Zhumabek:2023wka} for the rest of the setup and notation. In particular, we use the present-day values $\Omega_{m0} =0.3$ and $\Omega_{r0}= 9\times 10^{-5}$. 
We adopt the setup of~\cite{Zhumabek:2023wka}, where the Exp II model was referred to as ExpLin. Here we use it for the updated Exp potentials considered here.

\section{Dark energy of Exp II in the CPL parametrization with $w_0, w_a$}

We begin with the Exp II potential \cite{Dimopoulos:2017zvq}
\begin{equation}
    V(\vp)
    = M^2 e^{-2g}  \Bigl(e^{g \bigl( \tanh \frac{\vp}{\sqrt{6\alpha}} +1\bigr)}-1\Bigr)
    \label{ExpII}
\end{equation}
  and use it as the baseline model for the numerical analysis.
The starting point is the late-time scalar-field dynamics studied in detail in \cite{Akrami:2017cir,Dimopoulos:2017zvq,Akrami:2020zxw} and 
 \cite{Zhumabek:2023wka}.
 It was observed that in the CPL parametrization 
the model produces a thawing-like relation
\begin{equation}
    w_a \simeq -1.53(1+w_0).
    \label{eq:thawing_relation}
\end{equation}
We have checked that this relation is a robust prediction
of the Exp II potential, and  confirmed that it is largely unaffected by changes in the model
parameters, the initial conditions, or the small changes in the form of the scalar-field potential.

The values of  \(w_0\sim -0.8, \, w_a\sim -0.6\)   discovered by DESI \cite{DESI:2025zgx} are known to cross the ``phantom line''. The Exp II model of a single scalar field with a canonical kinetic term does not cross the line $w(z) =-1$ to reach $w(z) <  -1$.  Therefore, one can expect that no changes in the model
parameters, initial conditions, or the form of the scalar-field potential will lead to DESI values $ (w_0,w_a)$.

We  studied dark energy in the Exp II model using the CPL parametrization with $w_0, w_a$ following \cite{Zhumabek:2023wka}. 
The main scanned parameters are $\alpha$, the frozen initial field value $\varphi_f$, and the initial velocity variable $x_i=0$, which corresponds to a frozen thawing field at the initial time. In \cite{Zhumabek:2023wka}, the following range of parameters was investigated
\begin{equation}
    3\alpha = \{1,2,3,4,5,6,7\}, \qquad 
   \vp_f = \{-35,-15,-10,-9,-8\},
\end{equation}
with $x_i=0$.
We have checked that changing $\Omega_{\vp} $ among the values  $ [ 0.67, 0.7, 0.72 ]$ does not make much difference. Changing to smaller $\a$ mostly moves $(w_0,w_a)$ above the line defined by \eqn{eq:thawing_relation}. Changing the potential to a waterfall-modulated one does not change the dark energy evolution.  The closest values we found for  $(w_0,w_a)$ are approximately $(-0.8,\,-0.3) $.

\section{Evaluating the  equation of state   $w(z)$ for updated Exp models}
We can analyze dynamical dark energy by directly comparing the equation of state $w(z)$ with the data.  This was done, for example, for the axion-like quintessence potential $V(a)= m_a^2 f_a^2 (1+\cos {a/ f_a}) $ in \cite{DESI:2025fii}, where the result is presented in Fig. 12.
It was stressed there that the current data indicate
a clear preference for models that feature a phantom crossing. However, it was also noted that although alternative $w(z)$ parametrizations lacking this
feature are disfavored, they cannot yet be ruled out. 

An analogous comparison with the DESI data was performed in \cite{Jing:2026ymp} with regard to the Exp II quintessential $\a$-attractor model. The result for $w(z)$ was shown in Fig. 10 of \cite{Jing:2026ymp}. In this figure, the dark-energy $w(z)$  was shown  as a function of redshift $z$ for the best-fit $\a$-attractor Exp II model
(blue solid line), with the corresponding 1$\sigma$  confidence region (blue shaded band).

\begin{figure}[H]
\centering
\IfFileExists{DES_SUGRA_wz_expII_best_phi_bg_alpha_3_s.png}{%
\includegraphics[width=0.49\textwidth]{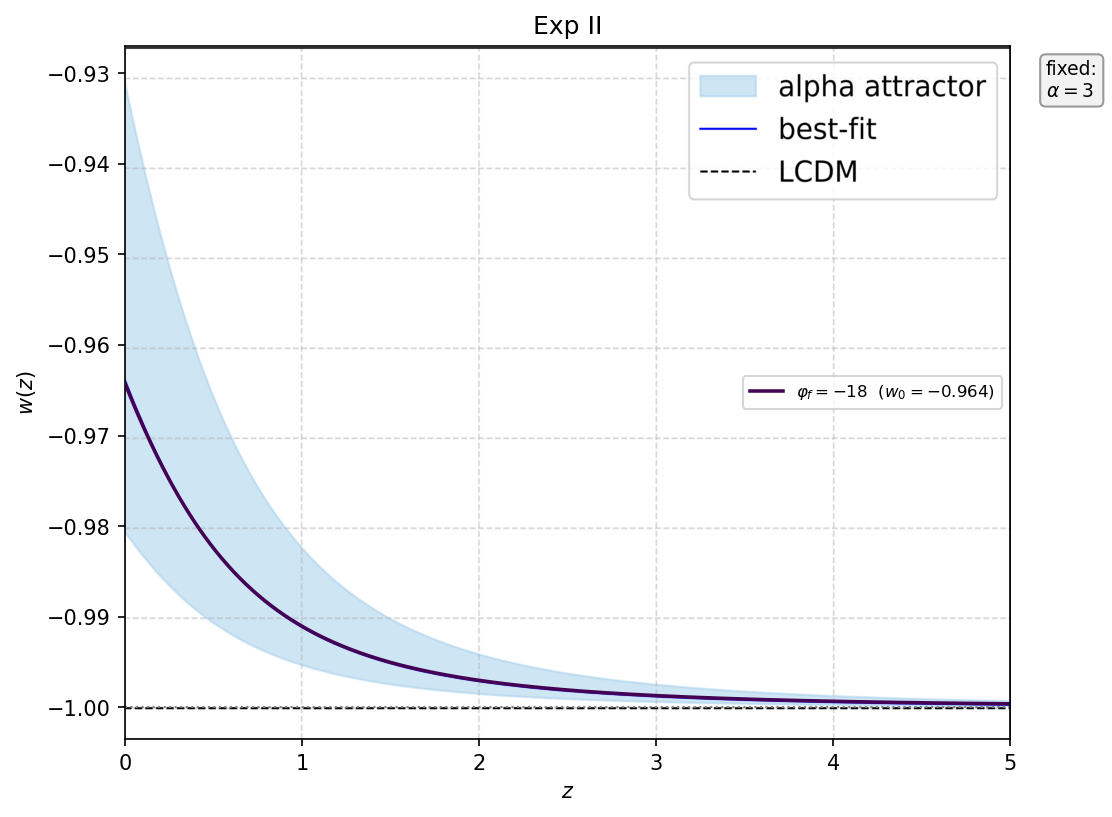}%
}{\fbox{\parbox[c][2.2in][c]{0.47\textwidth}{\centering Placeholder:
\texttt{..phi\_set\_bg\_alpha\_35.png} ($\alpha=5/3$).}}}\hfill
\IfFileExists{DES_SUGRA_wz_waterfall_alpha_3_phi_f_18_s.png}{%
\includegraphics[width=0.49\textwidth]{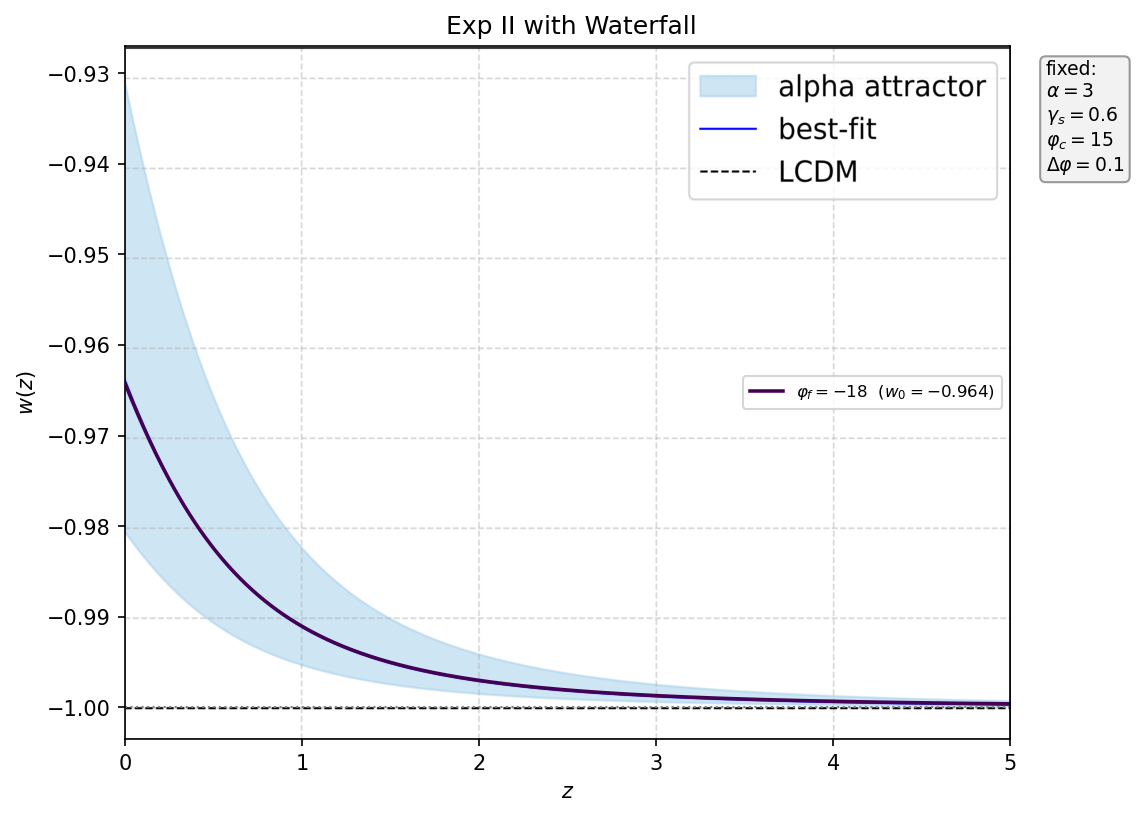}%
}{\fbox{\parbox[c][2.2in][c]{0.47\textwidth}{\centering Placeholder:
\texttt{..phi\_set\_bg\_alpha\_75.png} ($\alpha=7/3$).}}}
\caption{\footnotesize The light blue best-fit model in Fig. 10 of \cite{Jing:2026ymp}   is reproduced exactly here by the dark blue curve, representing the case $  \alpha=3$, $ \varphi_f=-18$. The left panel shows $w(z)$ for the model without the waterfall; the right panel shows $w(z)$ for the model with the waterfall. This clearly shows that the presence or absence of the waterfall feature during inflation does not affect dark energy.}
        \label{BestFit}
\end{figure}

\begin{figure}[H]
\centering
\IfFileExists{DES_SUGRA_wz_explin_alpha_set_bg_18_s.png}{%
\includegraphics[width=0.49\textwidth]{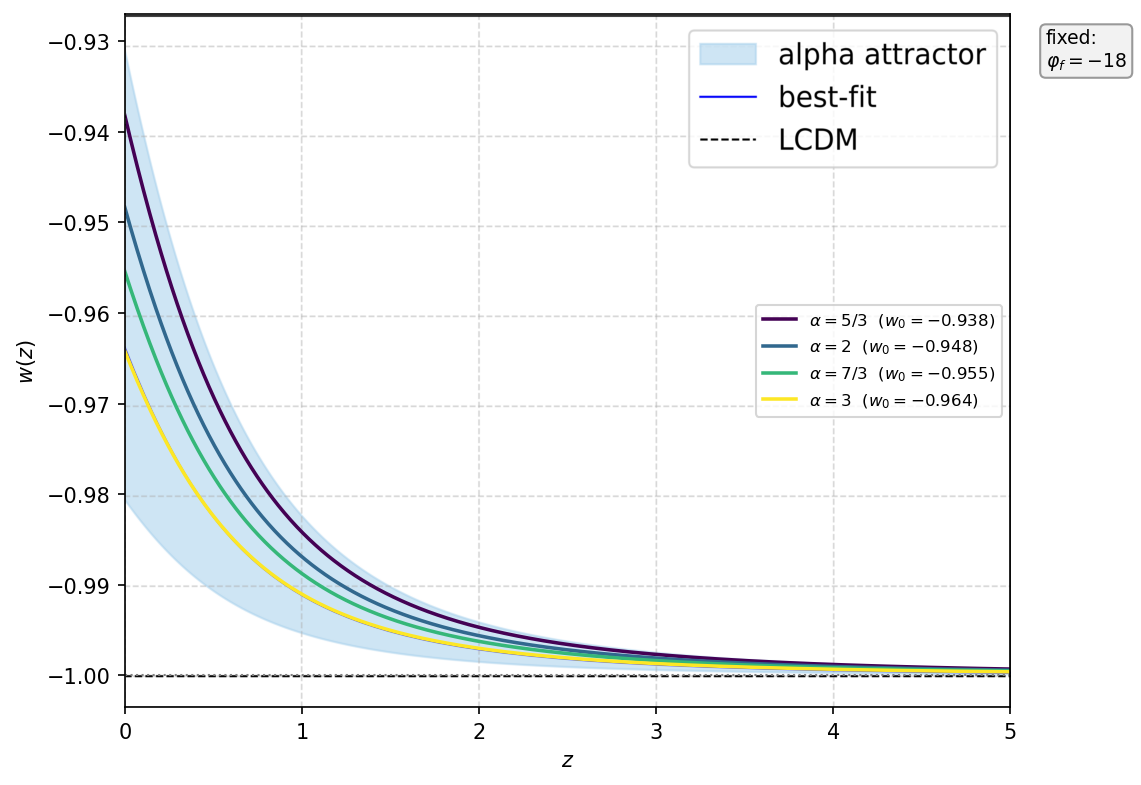}%
}{\fbox{\parbox[c][2.2in][c]{0.47\textwidth}{\centering Placeholder:
\texttt{..alpha\_set\_bg\_18.png} ($\phi_f=-18$).}}}\hfill
\IfFileExists{DES_SUGRA_wz_explin_alpha_set_bg_35_s.png}{%
\includegraphics[width=0.49\textwidth]{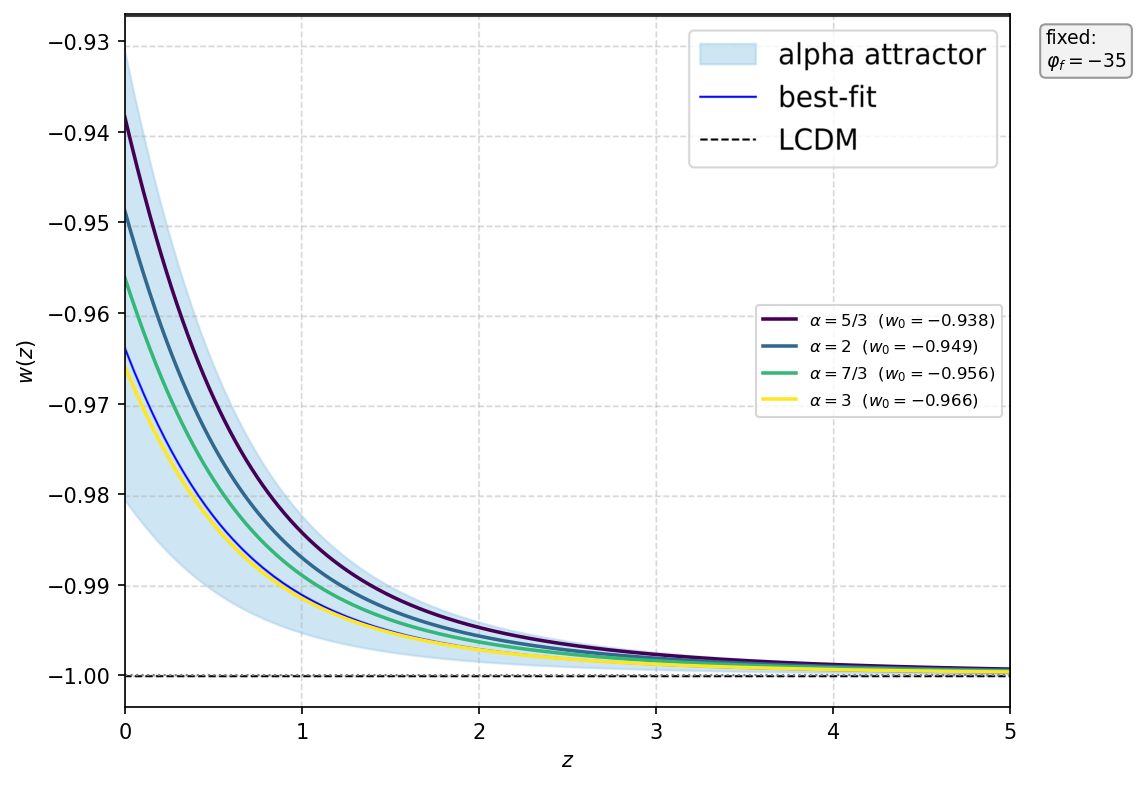}%
}{\fbox{\parbox[c][2.2in][c]{0.47\textwidth}{\centering Placeholder:
\texttt{..alpha\_set\_bg\_35.png} ($\phi_f=-35$).}}}
\caption{\footnotesize Model predictions for $w(z)$ on the same band for $\vp_f=-18$ (\emph{left}) and $\vp_f=-35$
(\emph{right}); within each panel the curves span several values of $\alpha$.  Varying
$\alpha$ changes the attractor tail slope $2/\sqrt{6\alpha}$ and the amplitude of the
thaw.}
\label{Smaller}
\end{figure}

\begin{figure}[H]
\centering
\IfFileExists{DES_SUGRA_wz_delta_DS_a53_s.png}{%
\includegraphics[width=0.49\textwidth]{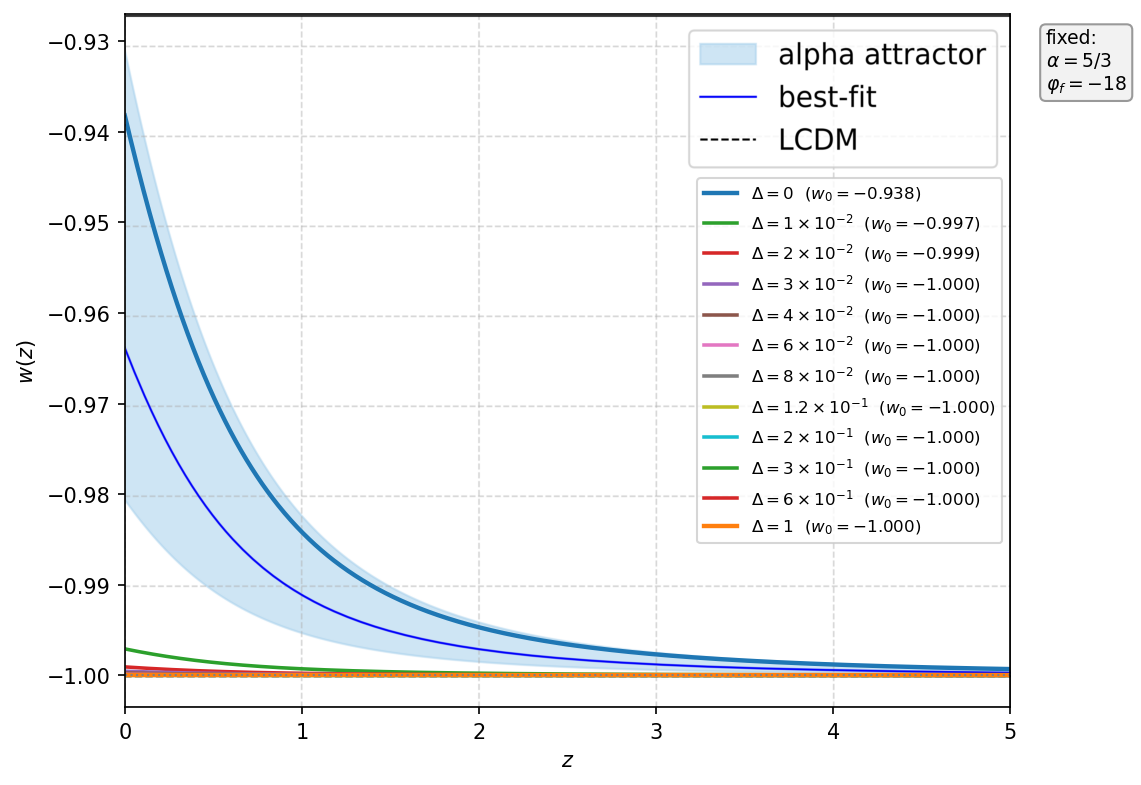}%
}{\fbox{\parbox[c][2.0in][c]{0.47\textwidth}{\centering Placeholder:
\texttt{figures/DES\_SUGRA\_wz\_k\_a53.png} ($\alpha=5/3$, $\phi_f=-18$, $\Delta=0\!\to\!1$).}}}\hfill
\IfFileExists{DES_SUGRA_wz_delta_DS_a2_s.png}{%
\includegraphics[width=0.49\textwidth]{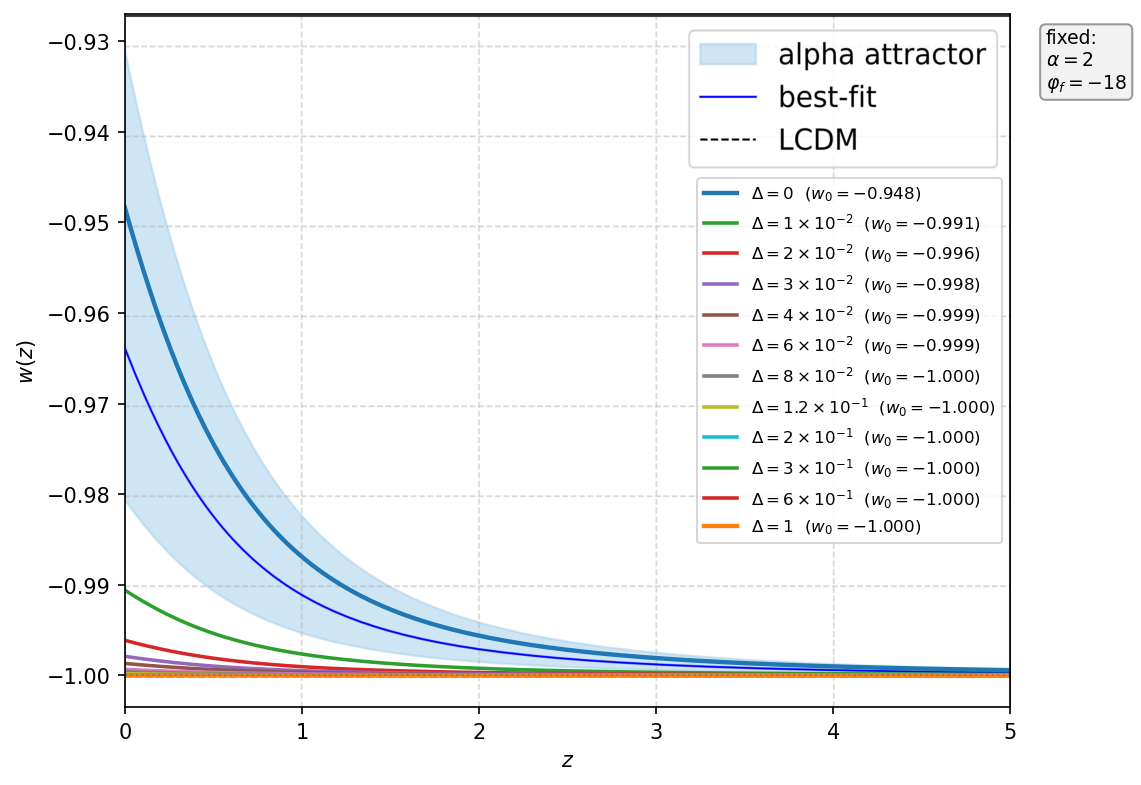}%
}{\fbox{\parbox[c][2.0in][c]{0.47\textwidth}{\centering Placeholder:
\texttt{figures/wz\_k\_a2.png} ($\alpha=2$, $\phi_f=-18$, $\Delta=0\!\to\!1$).}}}

\vspace{0.4em}

\IfFileExists{DES_SUGRA_wz_delta_DS_a73_s.png}{%
\includegraphics[width=0.49\textwidth]{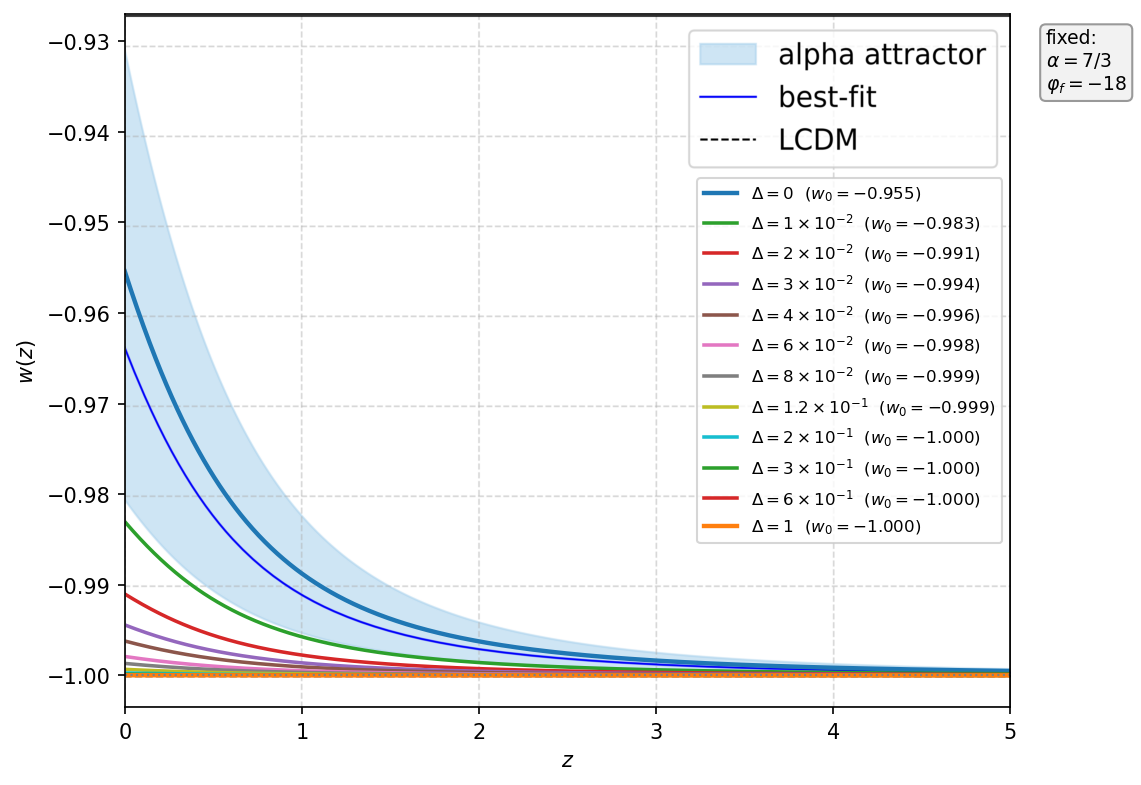}%
}{\fbox{\parbox[c][2.0in][c]{0.47\textwidth}{\centering Placeholder:
\texttt{figures/wz\_k\_a73.png} ($\alpha=7/3$, $\phi_f=-18$, $\Delta=0\!\to\!1$).}}}\hfill
\IfFileExists{DES_SUGRA_wz_delta_DS_a3_s.png}{%
\includegraphics[width=0.49\textwidth]{DES_SUGRA_wz_delta_DS_a3_s.png}%
}{\fbox{\parbox[c][2.0in][c]{0.47\textwidth}{\centering Placeholder:
\texttt{figures/wz\_k\_a3.png} ($\alpha=3$, $\phi_f=-18$, $\Delta=0\!\to\!1$).}}}
\caption{\footnotesize Equation-of-state  $w(z)$ for models interpolating from  Exp~I\,$\to$\,II, for
$\alpha= 5/3,\,2,\,7/3,\,3$ (\emph{top-left, top-right, bottom-left, bottom-right}) at fixed
$\vp_f=-18$; within each panel the curves span $\Delta=0$ (Exp~II, Minkowski tail) to
$\Delta=1$ (Exp~I, de~Sitter tail).  Colours match the $\Delta$ ordering of the potentials in Fig.~\ref{InterPot}.}
\label{Interpolating}
\end{figure}

Here we numerically evaluate the function $w(z)$ for the Exp quintessential $\a$-attractor models and superimpose the resulting curves on the DESI data given in Fig. 10 of \cite{Jing:2026ymp}. First, we find an explicit model that fits the best-fit line in Fig. 10 of \cite{Jing:2026ymp}. We show this in  Fig. \ref{BestFit}. The best-fit Exp II model had $\a=3$ and $\vp_f=-18$, both with and without the waterfall, allowing higher $n_s$ in these quintessential models.
Figure~\ref{Smaller} 
fixes $\vp_f$ ($-18$ and $-35$) and varies $\alpha$: for smaller $\a<3$, all $w(z)$ curves tend to be above the best-fit light blue curve for both values of $\vp_f$ ($-18$ and $-35$).

In Fig. \ref{Interpolating}, we show the $w(z)$ curves for interpolating de Sitter-to-Minkowski models with $\Delta$ changing as shown in \eqn{CCflex}. Our choice of $\Delta$ values was motivated by the desire to have a dense set of interpolating curves in the bottom-right panel of Fig. \ref{Interpolating} for $w(z)$.

Since $w_0$ runs from its Exp~II value down to $-1$ over the narrow interval $\Delta\lesssim g\,(1+\tanh(\vp_f/\sqrt{6\a}))\sim10^{-2}$, the values of $\Delta$ in \eqn{CCflex} are spaced non-uniformly, more densely toward $\Delta=0$, so that the transition is clearly defined.

Thus, we learn here that interpolating models cover the light-blue 1$\sigma$ region of DESI, and also go down to Exp I, which is a model with maximal cosmological constant $\Lambda= M^2 e^{-2g}$ in this class of models.
These models are ready to be tested by upcoming dark energy experiments, particularly if they are closer to  $\Lambda$CDM than DESI results suggest.

\section{Inflaton-axion quintessential $\a$-attractor  scenario}\label{Sec:axion}

We start with the Exp II model in \eqn{ExpII}, which at $\vp\to -\infty$ exponentially tends to a Minkowski universe with $V\to 0$  
\be
V_{\vp\to -\infty} \to  2gM^2 e^{-2g}  e^{\sqrt{2\over 3\a}  \vp}\Big| _{\vp\to -\infty}  \to 2gM^2 e^{-2g}  e^{-\sqrt{2\over 3\a}  |\vp|} \to 0\, .
\label{tend}\ee
We studied the evolution of dark energy in the single-field Exp II model in the earlier sections. 
Now we generalize this model to include the axion, to match the quintessence model in \cite{Toomey:2025yuy,Chudaykin:2026amr}.

There is a subtlety in comparing our quintessential model, with inflation at positive $\vp$ and evolving dark energy at negative $\vp$,  with the quintessence model in \cite{Toomey:2025yuy}, where the field $\chi$ is non-negative and rolls  to positive infinity in the future. Our field $\vp$ is positive during inflation and negative during the evolution of dark energy; see, for example, Fig. \ref{PotQ} here. Therefore, 
the field $\chi$ in \cite{Toomey:2025yuy}, which is  positive,  corresponds to negative $\vp$ in the quintessential model. This means that in making this comparison, one should remember that the quintessence field $\chi$ is related to the quintessential field $\vp$ as 
\be
\chi\to-\vp \qquad \text{for }\vp<0\, .
\ee
This subtlety is important in identifying the axion field metric; namely, in \cite{Toomey:2025yuy} we have
\be
e^{\lambda \chi} (\partial a)^2
\ee
with negative $\lambda$. In hyperbolic geometry, we have two choices
\be
e^{\mp \, 2\, \mu \, \vp}  (\partial a )^2\, 
\ee
To match \cite{Toomey:2025yuy}, we take into account that 
\be
e^{\lambda \chi} \to e^{-| \lambda |  \chi} = e^{|\lambda |  \vp} \ ,
\ee
where we have used the fact that $\lambda=-|\lambda|<0$.
The inflaton-axion models of $\a$-attractors were studied in detail in \cite{Achucarro:2017ing,Yamada:2018nsk,Linde:2018hmx,Kallosh:2022vha,Carrasco:2025rud}. 
In hyperbolic geometry with the \K curvature $
\mathcal{R}_K= - {2\over 3\a} 
$, the $SL(2, \mathbb{R})$-invariant kinetic term depends on a half-plane geometric variable ${\rm Re}\,  T > 0$
\be
-3\a {\partial T \partial \bar T\over (T+\bar T)^2}\, .
\label{hyper}\ee
One can switch to a canonical variable $\vp$ 
\be
T= e^{\pm \mu\,  \vp} + i \mu \, a, \qquad \mu= \sqrt{2\over 3\a} = \sqrt{-\mathcal{R}_K} >0\, .
\ee
The  kinetic term in \eqn{hyper} acquires the form 
\be
-{1\over 2} [(\partial \vp )^2 + e^{\mp \, 2\, \mu \, \vp}  (\partial a )^2] \ ,
\ee
where the axion field has a $\vp$-dependent metric $g_{aa}=e^{\mp \, 2\, \mu \, \vp} $. There are two possibilities that have been discussed in the past:

1. The sign of the exponent in the metric is positive, $g_{aa}=e^{ \, 2\, \mu \, \vp} $. Inflationary models of this kind, where both $\vp$ and $a$ might evolve during inflation, were studied in \cite{Achucarro:2017ing,Yamada:2018nsk,Linde:2018hmx,Kallosh:2022vha}. In such a case, a phenomenon called ``Universality of multi-field
$\a$-attractors'' was discovered. The effect of this geometry was to impose friction on the axion evolution at large $\vp$. As a result, in these models, where inflation takes place near the half-plane boundary, at ${\rm Re}\,  T  \to  0$, the axion-inflaton model was effectively a model of a single field $\vp$ which evolves during inflation. We also note that the large effective decay constant can suppress the isocurvature perturbations of the axion during inflation.

2. An opposite sign in the exponent in the metric $g_{aa}=e^{ -\, 2\, \mu \, \vp} $ was encountered in $SL(2, \mathbb{Z})$-invariant modular cosmology models where a procedure for axion stabilization during inflation at large $\vp$ was implemented in \cite{Carrasco:2025rud}. Axion stabilization during inflation led to a single-field inflationary model. The axion field in $SL(2, \mathbb{Z})$ models starts moving only at small $\vp$, after inflation, where the effect of stabilization gradually disappears.

The $\a$-attractor quintessence axion-dilaton 
 model in  \cite{Toomey:2025yuy,Chudaykin:2026amr} fits the DESI data while preserving the null energy condition and never crossing the phantom line. It explains how, in single-field quintessence models with a canonical kinetic term, the same data would suggest phantom-line crossing. In the axion-dilaton quintessence model in \cite{Toomey:2025yuy,Chudaykin:2026amr},
the kinetic term is \footnote{We are grateful to M. Ivanov for explaining that   the sign in the exponent in $g_{aa}$  has to be negative in terms of the field $\chi$ in order to reproduce the ``effective phantom crossing.'' }
\be
-{1\over 2} [(\partial \vp )^2 + e^{2 \mu \, \vp}  (\partial a )^2]|_{\vp< 0} \quad \to \quad  -{1\over 2} [(\partial \chi )^2 + e^{-2 \mu \, \chi}  (\partial a )^2],\qquad \lambda = -2\mu = -2\sqrt{2\over 3\a} \,.
\ee
The choice of the sign of the exponent in $g_{aa}$, in contrast to the cases studied in   \cite{Achucarro:2017ing,Yamada:2018nsk,Linde:2018hmx,Kallosh:2022vha}, where the choice was a priori optional and decided by the model builders, was determined by the DESI data in  \cite{Toomey:2025yuy,Chudaykin:2026amr}.

Thus, to have the quintessence case of  \cite{Toomey:2025yuy,Chudaykin:2026amr} in the dark energy part of the quintessential model, our choice of the metric is as in case 1 above, when the sign of the exponent in the metric is positive: $g_{aa}=e^{ \, 2\, \mu \, \vp} $.  Therefore, during inflation, we  have   ``Universality of multi-field
$\a$-attractors'': The axion does not move since the effective derivative of the potential with respect to the axion field is strongly suppressed by the factor coming from the inverse metric, $e^{-2 \mu \, \vp} \, {\partial V\over \partial a} $.  This is the  ``rolling on the ridge'' effect \cite{Achucarro:2017ing}.   Meanwhile, during the dark energy stage, the same factor becomes $e^{2 \mu \, |\vp|} \, {\partial V\over \partial a} $ and it pushes the axion away from the ridge, as the data require \cite{Toomey:2025yuy}.

The quintessence potential in Eq. (5) of \cite{Toomey:2025yuy}, written in our notation (with $\beta$ replacing $\a$ and $V_0^q$ replacing $V_0$, since we use the parameters $\a$ and $V_0$ differently in this paper), is 
\be
V(\vp, a) \to V(\chi,a)= V_0^q e^{-\beta \chi} + m^2 \, f_a^2 [1-\cos(a/f_a)]\,,
\label{Mpot}\ee
where we have used $\varphi\to -\chi$.
The first term in \eqn{Mpot}  is already present in our single-field models; see    \eqn{tend}, so that 
\be
V^q_0= 2gM^2 e^{-2g} \, , \qquad \beta = \sqrt{2\over 3\a} \, .
\ee
Now we have to add to the potential in \eqn{ExpII} the axion-dependent term of the form
\be
m^2 f_a^2 [1-\cos(a/f_a)]\, .
\label{axion}\ee
Due to the dynamical stabilization of the axion field for $\vp > 0$ \cite{Achucarro:2017ing,Yamada:2018nsk}, the quintessential potential during inflation is not required to stabilize the axion in this regime, since the axion does not move anyway. However, when the field $\vp$ becomes small, the axion may actually move, as we have shown in \cite{Achucarro:2017ing,Yamada:2018nsk,Linde:2018hmx,Kallosh:2022vha}.  To construct a consistent quintessential axion-dilaton $ \alpha$-attractor model, we would like the axion $a$ to be destabilized at the value of the field $\varphi$ required for a successful quintessence model in \cite{Toomey:2025yuy}, compatible with DESI.  

Closely related models were already constructed in \cite{Linde:2018hmx} with a $\vp$-dependent axion destabilization point. In such a case, one might consider the quintessential model  with the following kinetic and potential terms
\be
-{1\over 2} [(\partial \vp )^2 + e^{2\sqrt{2\over 3\a} \, \vp}  (\partial a )^2] \ ,
\ee
\be
 V^{\rm quint}(\vp, a)
    = M^2 e^{-2g}  \Bigl(e^{g \bigl( \tanh \frac{\vp-\vp_{0}}{\sqrt{6\alpha}} +1\bigr)}-1\Bigr) + m^2 f_a^2 [1-\cos(a/f_a)] \ .
    \label{shift}
\end{equation}
 In this model, destabilization begins when $\vp$ becomes negative. But the shift of the potential by $\vp_{0}$ allows us to set the beginning of the dark energy stage at any desired point $\vp = \vp_{f}$ where kination stops and the field $\vp$ temporarily freezes. In particular, we may set $\vp_{f}$ close to the value of the field $\vp$ where both fields begin to move, as in the model \cite{Toomey:2025yuy}.  This could provide a continuous transition from the effectively single-field quintessential inflation in the theory \rf{shift} 
 with the axion frozen for $\vp \gg 0$  to the quintessential inflaton-axion evolution at $\vp < \vp_{f}$ as described in   \cite{Toomey:2025yuy}. 
 
 At present, this remains a plausible scenario that needs further development, clarification, and improvement, particularly if new experiments such as Euclid and Rubin support the DESI dark energy data in the near future.  One may hope that the tools developed in~\cite{Achucarro:2017ing,Yamada:2018nsk,Linde:2018hmx,Kallosh:2022vha} will help with a more detailed investigation of this model.

\section{Reheating in quintessential $\a$-attractors}\label{Sec: rehe} 

In quintessential inflation, the inflaton does not oscillate around a minimum after inflation, and conventional perturbative reheating through inflaton decay is therefore unavailable. Instead, the transition from inflation to kination changes the expansion law nonadiabatically and gravitationally produces non-conformally invariant fields~\cite {Parker:1968mv,Parker:1969au,Parker:1971pt,Zeldovich:1971mw,Kolb:2023ydq}. This mechanism requires no direct inflaton couplings, which helps preserve the approximate shift symmetry. Furthermore, the waterfall modulation allows the inflationary observables to mimic those associated with a larger effective plateau e-fold number, without requiring a prolonged kination epoch to increase the physical value of $N_*$.

For an order-of-magnitude estimate, we approximate the transition as instantaneous and write the energy density produced in each real, effectively massless, non-conformal bosonic degree of freedom as
\begin{equation}
\rho_{i,{\rm end}}\simeq cH_{\rm end}^{4},\qquad c\sim10^{-2},
\label{eq:grav-production-density}
\end{equation}
where $c$ depends on the duration and smoothness of the transition~\cite{Chun:2009yu}. In the minimal setup, the four Higgs degrees of freedom form the visible component, while the two graviton polarizations and the inflaton--axion fluctuations give four dark-radiation degrees of freedom. For $H_{\rm end}\sim10^{13}\ {\rm GeV}$, this gives
\begin{equation}
T_{\rm reh}^{\rm min}\simeq1.2\times10^{6}\ {\rm GeV}
\left(\frac{c}{10^{-2}}\right)^{3/4}
\left(\frac{H_{\rm end}}{10^{13}\ {\rm GeV}}\right)^2,
\qquad
\Delta N_{\rm eff}^{\rm min}\simeq2.9.
\label{eq:minimal-reheating-summary}
\end{equation}
Thus the temperature is safely above the BBN scale, but the dark-radiation contribution exceeds the conservative bound $\Delta N_{\rm eff}\lesssim0.3$~\cite{Planck:2018vyg}. Even counting only gravitons gives $\Delta N_{\rm eff}\simeq1.4$. This tension is a problem for the minimal field content, rather than a generic failure of gravitational reheating.

%\subsection{Enhancing visible-sector production}

One may reduce the energy density in gravitational radiation by using the mechanism of instant preheating \cite{Felder:1998vq,Felder:1999pv,Kofman:2004yc,Akrami:2017cir,Dimopoulos:2017tud}.
One may add to the original theory a field $\sigma$ interacting with $\phi$ through ${g_{\sigma}^{2}\over 2}\phi^{2}\sigma^{2} = 3\alpha {g_{\sigma}^{2}} \sigma^{2 }\tanh^{2}{\vp\over \sqrt{6\alpha}}$ where $\phi$ is an original variable $\phi=\sqrt{6\alpha}\tanh \frac{\varphi}{\sqrt{6\alpha}}$ in T-models. This term stabilizes $\sigma$ at $\sigma = 0$ and therefore does not modify the inflaton potential. 
When the field $\phi$ moves through the point $\phi = 0$, the field $\sigma$ becomes massless, and then its mass grows again as $g_{\sigma}|\phi|$.  This creates a gas of $\sigma$ particles. At $|\vp| \ll \sqrt{6\alpha}$, the field $\phi$ and the canonically normalized inflaton $\vp$ coincide, so the theory of this process coincides with the theory developed in \cite{Felder:1998vq,Felder:1999pv}.  Thus we can use the results of \cite{Felder:1998vq,Felder:1999pv} for the energy density of the produced $\sigma$ particles:
\begin{equation}\label{supprrr}
 \rho_{\sigma}  \approx  {({g_{\sigma}|\dot\vp_0|})^{3/2} \over
8\pi^3}\left(\frac{a_0}{a(t)}\right)^3 \,  g_{\sigma} |\vp| \, ,
\end{equation}
 where $a_0$ denotes the scale factor at the time when $\sigma$ particles are produced. This energy density provides a contribution to the potential that grows in both directions away from $\vp = 0$. For sufficiently large $g_{\sigma}$, this may lead to a temporary trapping of the field $\vp$ near $\vp = 0$ \cite{Kofman:2004yc}. 

More generally, one may consider interaction terms $3\alpha g_{\sigma}^2 \sigma^{2 }\tanh^{2}{\vp- \vp_{c}\over \sqrt{6\alpha}}$, which may lead to instant preheating  at any desired point $\vp = \vp_{c}$ \cite{Kofman:2004yc}. In the context of quintessential inflation, it is more convenient to obtain such terms working in half-plane $T$-variables, where $ \tanh{\vp\over \sqrt{6\alpha}}= {2-T-\bar T\over 2+T+\bar T}$. The way to obtain terms of the type of $\tanh^{2}{\vp- \vp_{c}\over \sqrt{6\alpha}}$ in this context is explained in our recent paper \cite{Kallosh:2026kfx}.

  The energy density of massive particles created in this process decreases more slowly than the energy density of gravitational radiation, which reduces the relative contribution of gravitational radiation to the total energy density.
  
  A detailed analysis of instant preheating in Exp II quintessential $\a$-attractors was performed in \cite{Jing:2026ymp} with the same conclusion that this mechanism can bring the gravitational-wave contribution to $\Delta N_{\rm eff}$ within the observational bound. However, this mechanism shortens the kination stage in this scenario, so one cannot increase $n_s$ along the lines of \cite{Akrami:2017cir,Dimopoulos:2017tud}. Therefore, there was a remaining problem with quintessential $\a$-attractors due to a small value of $n_s$, incompatible with ACT+DESI. But in the new scenario proposed in this paper, the $n_s$  problem is solved since the waterfall insertion can easily increase $n_{s}$. Thus, in our updated models, we do not need a prolonged kination stage to raise $n_s$.

Another way to reduce the dark-radiation fraction is to increase the number of visible non-conformal fields. In the MSSM, the sfermions and two Higgs doublets provide $N_{\rm vis}^{\rm MSSM}=98$ real scalar degrees of freedom~\cite{Martin:1997ns}. If their effective masses during the transition are much smaller than $H_{\rm end}$, gravitational production is efficient. For $m_{\rm SUSY}\sim10^{7}\ {\rm GeV}$ and $H_{\rm end}\sim10^{13}\ {\rm GeV}$, the produced superparticles remain relativistic through most of kination. Assuming $R$-parity violation, they eventually decay into Standard Model particles.\footnote{The observed dark matter abundance is assumed to arise from another candidate outside the LSP sector.} The resulting estimates are
\begin{equation}
T_{\rm reh}^{\rm MSSM}\simeq(0.8\mathord{-}1.0)\times10^{7}\ {\rm GeV}
\left(\frac{c}{10^{-2}}\right)^{3/4}
\left(\frac{H_{\rm end}}{10^{13}\ {\rm GeV}}\right)^2,
\qquad
\Delta N_{\rm eff}^{\rm MSSM}\simeq0.09\mathord{-}0.12.
\label{eq:mssm-reheating-summary}
\end{equation}
These values satisfy the dark-radiation bound. For a representative two-body decay, decay before radiation--kination equality requires approximately $\lambda_{\rm RPV}\gtrsim2\times10^{-5}$ at $m_{\rm SUSY}=10^{7}\ {\rm GeV}$; decay before BBN is considerably easier. Additional visible-sector entropy release would further reduce $\Delta N_{\rm eff}$.

One may also reduce the relative amount of dark radiation by considering a positive Higgs-curvature coupling
\begin{equation}
\mathcal{L}_{H}\supset-\xi RH^{\dagger}H,
\end{equation}
with $\xi>0$. This makes the Higgs field heavy during inflation but tachyonic after the transition to kination, thereby amplifying Higgs fluctuations and preferentially reheating the visible sector~\cite{Nakama:2018gll} (see also \cite{Opferkuch:2019zbd} for a related work).  For the representative choice $\xi\simeq2$, the analysis of Ref.~\cite{Nakama:2018gll} gives
\begin{equation}
T_{\rm reh}^{(H)}\simeq(1.5\text{--}1.9)\times10^{7}\ {\rm GeV}
\left(\frac{H_{\rm end}}{10^{13}\ {\rm GeV}}\right)^2,
\qquad
\Delta N_{\rm eff}^{(H)}\simeq0.13\text{--}0.18,
\label{eq:higgs-reheating-summary}
\end{equation}
where the dark-radiation estimate above conservatively includes gravitons and the inflaton--axion fluctuations, which are not included in the setup of \cite{Nakama:2018gll}. The result remains below $0.3$, although its precise value depends on the transition profile and the running of the Higgs quartic coupling.

Another non-supersymmetric possibility is the gravitational production of massive species $X$ followed by its decay into the visible sector~\cite{Hashiba:2018grh,Hashiba:2018gpc,Fujikura:2022gpp}. Once non-relativistic, $\rho_X\propto a^{-3}$ grows relative to both kination, $\rho_{\rm kin}\propto a^{-6}$, and dark radiation, $\rho_{\rm DR}\propto a^{-4}$. A temporary $X$-dominated epoch therefore suppresses $\rho_{\rm DR}/\rho_X$, and the subsequent decay injects entropy into the visible sector that further dilutes $\Delta N_{\rm eff}$. The efficiency is model-dependent, but the waterfall mechanism makes an extended kination epoch unnecessary in this construction.

The minimal gravitational reheating scenario therefore has a dark-radiation problem, but it is not the only possibility. Using the instant preheating mechanism, enlarging the visible field content, tachyonically amplifying Higgs fluctuations through its curvature coupling, or introducing a decaying heavy species can all yield reheating temperatures above the BBN scale while satisfying the dark-radiation bound.  Thus, several possible reheating scenarios can make waterfall-updated quintessential $\a$-attractors compatible with the current experimental data, but we leave further details for future work.  

\section{Dynamical dark energy and the fate of the universe }\label{Sec:fate} 
 
We have studied various dark energy models with a negative cosmological constant (CC) in~\cite{Kallosh:2002gf, Kallosh:2002gg, Kallosh:2003mt, Gutperle:2003kc, Kallosh:2003bq}. They predict that the universe will collapse at some point in the future. 

Here we study the exponential dark-energy models introduced in Sec. \ref{Sec:Exp}, which have positive, vanishing, and negative CC $\Lambda = \Delta M^2 e^{-2g}$, where 
\be
\Delta= (1, \, 0.02, \, 0, \, - 0.01, \, - 0.02 ) 
\ee

%Thus $k=1$ (the observational best-fit model of Ref.~\cite{Jing:2026}) is the exact
%\emph{watershed} between an eternally expanding and a collapsing universe, and any
%$k>1$ --- an arbitrarily small extra negative shift of the SUGRA vacuum energy --- makes
%the potential AdS in the tail and the universe collapse.

\begin{figure}[H]
\centering
\IfFileExists{DES_EXP_FATE_4_wz.png}{%
\includegraphics[width=0.5\textwidth]{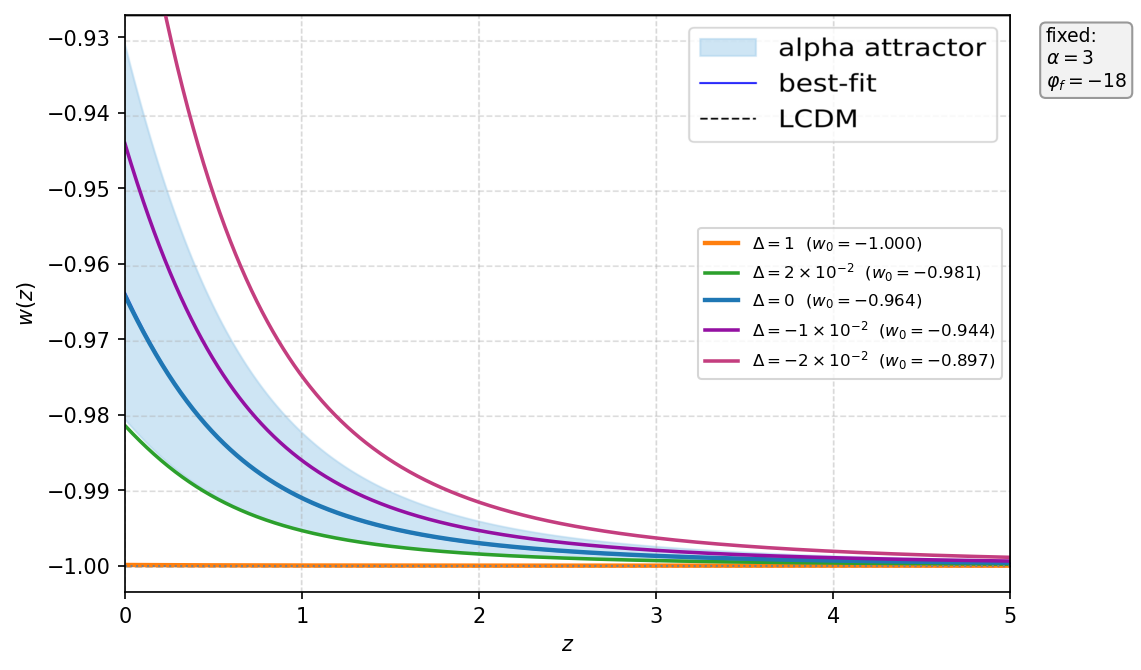}%
}{\fbox{\parbox[c][2.2in][c]{0.46\textwidth}{\centering Placeholder:
\texttt{..alpha\_set\_bg\_18.png} ($\phi_f=-18$).}}}\hfill
\IfFileExists{DES_EXP_FATE_4_at.png}{%
\includegraphics[width=0.49\textwidth]{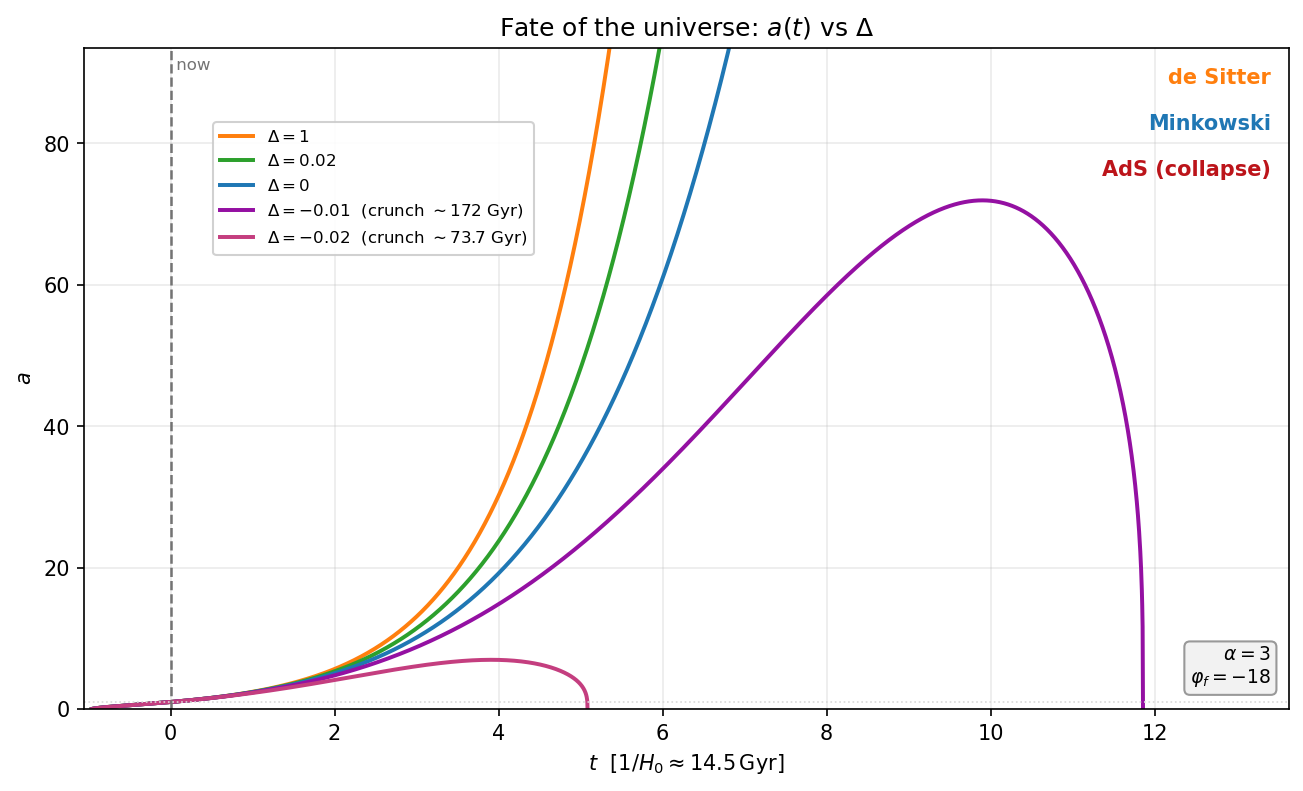}%
}{\fbox{\parbox[c][2.2in][c]{0.47\textwidth}{\centering Placeholder:
\texttt{..alpha\_set\_bg\_35.png} ($\phi_f=-35$).}}}
\caption{\footnotesize $w(z)$  (\emph{left})  and $a(t)$  (\emph{right}) for de Sitter, Minkowski, and collapsing AdS universes in Exp models with $\alpha = 3$ and  $\vp_f=-18$.  All lines in the left panel in the area below the blue curve ($\Lambda = 0$) correspond to future dS universes with a positive cosmological constant. All lines in the area above the blue curve ($\Lambda = 0$) correspond to a negative cosmological constant and a future collapse.}
\label{ds_m_ads}
\end{figure}
The collapsing branch can be added to the Exp interpolation family
Eq.~\eqref{CCflex}: pushing the cosmological constant below $\Lambda = 0$
turns eternal expansion into a future collapse.  
\begin{equation}
    \Lambda =\Delta \, M^2e^{-2g}
    \ \ \begin{cases}>0 & \Delta>0\ \ (\text{dS, Exp~I $\to$  II}),\\
                     =0 & \Delta=0\ \ (\text{Minkowski, Exp~II}),\\
                     <0 & \Delta< 0\ \ (\text{AdS,  collapse}).\end{cases}
    \label{eq:explinR-tail}
\end{equation}
In Fig. \ref{ds_m_ads} we show the equation of state $w(z)$ for Exp models with $\Lambda$ positive, vanishing, and negative. The cases with negative $\Lambda$ lie above those with positive or vanishing values, which is the signature of collapsing universes. We also give the values of $a(t)$ for all these models. It is clear that those with negative $\Lambda$ are collapsing universes. Those with larger $|\Lambda|$ have a shorter lifetime. 
\begin{figure}[H]
\vskip 0.3cm 
\centering
		 \includegraphics[width=0.9\textwidth]{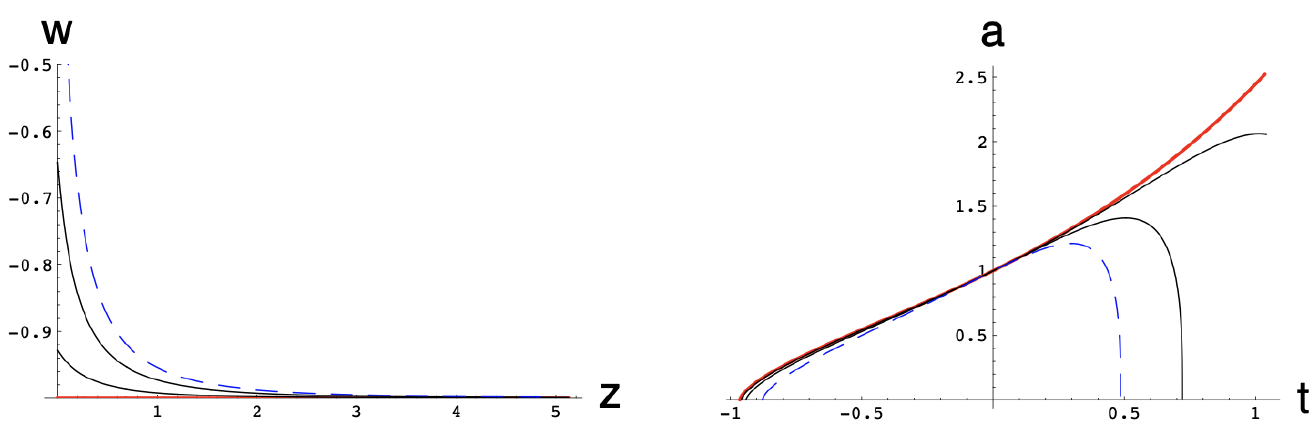}
        \caption{\footnotesize  The gauged supergravity N=8 potential studied in \cite{Kallosh:2002gf} is shown for different deviations of the initial point of the canonical scalar field from the de Sitter maximum of the potential $V= \Lambda (2- \cosh \sqrt 2 \, \vp)$. We plot $w(z)$ and the scale factor $a(t)$. All initial conditions away from the maximum result in a collapse of the universe. 
        } 
        \label{PotQ3}
\end{figure}
The qualitative feature of the plots in Fig. \ref{ds_m_ads} is reminiscent of the case of the gauged $N=8$ supergravity dark-energy potential studied in \cite{Kallosh:2002gf}, where we plotted 
 $w(z)$  and $a(t)$. One can see in Fig. \ref{PotQ3} that all collapsing models have $w(z)$ above the single ``eternal'' one, in red.  
 
Cosmic-time evolution with negative potentials,  $V<0$, requires a careful approach and second-order dynamical equations. The autonomous variables of Sec.~\ref{Sec:evol} use
$y=\sqrt{V}/(\sqrt3 H)$ and are restricted to $V>0$.  To follow the field
into a negative-potential region, through the turnaround point ($H=0$) and the ensuing
collapse, we integrate the second-order system in time $t$ (units
$H_0^{-1}\simeq14.5\,$Gyr; $M_{\rm Pl}=1$),
\begin{align}
    \dot a &= \tilde a, &
    \dot{\tilde a} &= -\frac{a}{6}\left(\rho+3p\right), &
    \ddot\vp &= -3\frac{\tilde a}{a}\,\dot\vp - V'(\vp),
    \label{eq:raychaudhuri}
\end{align}
with $\rho+3p=\rho_m+2\rho_r+2\dot\vp^2-2V(\vp)$,
$\rho_m=3\Omega_{m0}H_0^2\,a^{-3}$ and $\rho_r=3\Omega_{r0}H_0^2\,a^{-4}$.  The second of
Eqs.~\eqref{eq:raychaudhuri} is the Friedmann--Lema\^itre form of an evolution equation for $\ddot a$ that contains
neither $H^{-1}$ nor $\sqrt{\rho}$, and is therefore regular for \emph{any} sign of $H$,
$\rho$ and $V$. The Friedmann constraint itself remains regular at the turnaround, where $H=\rho=0$. However, the expanding-branch formulation $H=+\sqrt{\rho/3}$ and the autonomous variables of Sec.~\ref{Sec:evol} do not provide a smooth continuation to $H<0$. The second-order system~\eqref{eq:raychaudhuri}, by contrast, passes smoothly through the turnaround. The Friedmann equation is imposed
only on the initial data and monitored thereafter as an accuracy check.  We start at the
present epoch ($t=0$, $a=1$, $\tilde a=H_0$) with the field at its thawing value
$\vp_0$ and velocity $\dot\vp_0$ taken from the late-time attractor solution of
Sec.~\ref{Sec:evol}, and fix the overall potential scale by
$\frac{1}{2}\dot{\vp}_0^2+V(\vp_0)=3H_0^2\Omega_{\vp,0}$ with $\Omega_{\vp,0}\simeq 0.7$.

The times to collapse are given in the table.
\begin{table}[H]
\centering
\begin{tabular}{cccccc}
\hline\hline
$\Delta$ & $w_0$ & turnaround & full collapse & $a_{\max}$ & fate \\
\hline
$1$ (Exp~I)   & $-1.000$ & --- & --- & $\infty$ & eternal dS \\
$0.02$           & $-0.981$ & --- & --- & $\infty$ & de Sitter (small plateau) \\
$0$ (Exp~II)  & $-0.964$ & --- & --- & $\infty$ & Minkowski (marginal) \\
$-0.01$ & $-0.944$ & $\sim143$ Gyr & $\sim172$ Gyr & $\sim72$ & slow collapse \\
$-0.02$ & $-0.897$ & $\sim57$ Gyr  & $\sim74$ Gyr  & $\sim7$  & collapse \\
\hline\hline
\end{tabular}
\caption{Collapse of the Exp model ($\alpha=3$, $\vp_f=-18$) as the
interpolation constant $\Delta$ is lowered from $1$  to the Exp~II ($\Delta =0$)  boundary into
the AdS regime ($\Delta<0$).  Times measured from the present epoch, in Gyr
($H_0^{-1}\simeq14.5\,$Gyr).  }
\label{tab:exp-collapse}
\end{table}

In cases with smaller $\alpha$, we find an analogous pattern:  models with positive $\Lambda$ interpolate between Exp I with $\Delta=1$ and Exp II with $\Delta =0$. Models with negative $\Delta < 0$ and negative $\Lambda$ have $w(z)$ above the curve with $\Delta=0$: these are collapsing universes. Moreover, with smaller $\a$, even a tiny negative CC leads to a future collapse of the universe; a further decrease in $\Lambda$ would lead to a faster collapse, even before the present time, which would exclude such models.

 Our analysis suggests that if, in the future, we find $r$ and $n_s$ with good precision, we will also know $\a$ in our class of models. For any given $\a$  and $\vp_f$, we know $w(z)$ at $\Delta=0$; curves below it correspond to models with a future de Sitter state, whereas curves above it correspond to collapsing universes, and we can compute the lifetime of the universe before collapse, as we have shown in detail in the case of $\a=3$.

Thus, future data on evolving dark energy and inflationary values of $(n_s, r)$ in the context of the updated quintessential $\alpha$-attractors may allow us to predict the fate of the universe in these models.
\section{Summary}
In the future, new experiments, such as Euclid,  Vera Rubin Observatory,  Nancy
Grace Roman Space Telescope, and the full DESI and DES programs will release their data on dark energy. The data might confirm the DESI results or deviate from them; in particular, they may move closer to $\Lambda$CDM.

If future data move away from the DESI preference toward a cosmological constant, our general Exp models, interpolating between a de Sitter state at infinity with maximal $\Lambda= M^2 e^{-2g}$ and a Minkowski state with $\Lambda =0$, will be ready to be tested by the new data. They are described by the future cosmological constant $\Lambda=\Delta \cdot M^2 e^{-2g}$ with $0\leq \Delta\leq 1$.

The corresponding potentials are given in the right panel of Fig. \ref{InterPot} and the samples of the evolving 
 dark energy equation of state $w(z)$ are given in Figs. \ref{Inter} and \ref{Interpolating}. These models gradually deviate from $\Lambda$CDM toward evolving dark energy, up to the DESI best-fit $w(z)$.
 
 These updated single-field quintessential $\a$-attractor models, with waterfall-modulated potentials, predict the values of $n_s, r$ satisfying the relation $r\simeq 3\a(1-n_s)^2$, with flexible $n_s$ as we have shown in  \cite{Chudaykin:2026amr,Kallosh:2026kfx}. They are capable of having a small controllable deviation from the cosmological constant; therefore, $w(z)$ will be controlled by the parameters of these models. 
 
If, however, the new data support DESI, including the phantom crossing line, it will be more suitable to consider a two-field inflaton-axion quintessential inflationary model described briefly in Sec.  \ref{Sec:axion}. We have explored the possibility of combining the quintessence $\a$-attractor model in  \cite{Toomey:2025yuy,Chudaykin:2026amr} with the updated quintessential $\a$-attractor models. Combining them consistently will require a more detailed study.

In conclusion,  the general interpolating Exp models of quintessential $\a$-attractors were introduced in \cite{Akrami:2017cir}. We updated them in this paper to allow a flexible $n_s$ due to a waterfall-modulated stage of inflation, including values of $n_s\approx 0.973$ and higher.  We have also updated the original reheating mechanism proposed in \cite{Akrami:2017cir,Dimopoulos:2017zvq} so that the updated models do not violate the existing bound on $\Delta N_{\rm{eff}}$.

These models  predict the values of the dark-energy equation of state $w(z)$  interpolating between $\Lambda$CDM and DESI, depending on the choice of parameters in these models. They also have flexible  values of $(n_s, r)$ satisfying the relation $r\simeq 3\a(1-n_s)^2$ as shown in \cite{Chudaykin:2026amr,Kallosh:2026kfx}.  Future observations, including the search for B modes, can test the prediction $r\simeq 3\a(1-n_s)^2$.

We studied the fate of the universe in our Exp models with a negative CC. We have found an explicit signature of collapsing models in quintessential $ \alpha$-attractor models. It would be important to repeat this analysis if the value of $\a$ is found by future cosmological observations.

 \section*{Acknowledgments}
We are grateful to  Y. Akrami, G. Alestas, A. Chudaykin, M. Ivanov and  O. H. E. Philcox for discussions of dark energy in the context of $\a$-attractor quintessence and quintessential $\alpha$-attractors. RK,  AL, and MS are supported by the Leinweber Institute for Theoretical Physics at Stanford. RK and AL are supported by the NSF grant PHY-2310429.  YY is supported by IBS under the project code IBS-R018-Y3-2026-a00.

 \bibliographystyle{JHEP}
\bibliography{lindekalloshrefs}
\end{document}